\documentclass[a4paper,11pt]{article}
\pdfoutput=1 

\usepackage{jheppub} 

\usepackage[T1]{fontenc} 
\usepackage{bm}
\usepackage{amsmath}
\usepackage{amsfonts}
\usepackage{amssymb}
\usepackage{graphicx}%
\usepackage{amscd}
\usepackage{extarrows}
\usepackage[stable]{footmisc}

\title{\boldmath A title with some math: $x=1$}

\author[a,b]{Zhi Hu,}
\author[b,c]{Mulin Yan,}
\author[a,b]{Sen Hu,}

\affiliation[a]{School of Mathematics, University of Science and
Technology of China\\Hefei, 230026, China}
\affiliation[b]{Wu Wen Tsun Key Laboratory of Mathematics, Chinese Academy of Science\\Hefei, 230026, China}
\affiliation[c]{School of Physics, University of Science and
Technology of China\\Hefei, 230026, China}

\emailAdd{halfask@mail.ustc.edu.cn}
\emailAdd{mlyan@ustc.edu.cn}
\emailAdd{shu@ustc.edu.cn}

\abstract{In this paper we  present the two-parameter dynamics which is implied by the law of inertia in flat spacetime. A remarkable perception is that $(A)dS_4$ geometry may emerge from the two-parameter dynamics, which exhibits some phenomenon of dynamics/geometry correspondence. We also discuss the
Unruh effects within the context of two-parameter dynamics. In the last section we construct various invariant actions with respect to the broken symmetry groups.}

\begin{document}
\title {Two-Parameter Dynamics and Geometry}

\maketitle
\section{ $(A)dS_4$ Geometry from Two-Parameter Dynamics}
We start with a triple $(\mathbb{R}^4,\eta, B)$ for the Minkowski  metric $\eta=\textrm{diag}(-1,1,1,1)
$ and a non-degenerate symmetric bilinear form $B$. We take the following ansatz in terms of the Lorentz coordinates $\{x^0=ct, x^1,x^2,x^3\}$
\begin{align}
  B_{\mu\nu}=A_0(x)\eta_{\mu\nu}+A_1(x)\frac{\eta_{\mu\alpha}\eta_{\nu\beta}(x^\alpha x^\beta)^{d_1}}{l_1^{2d_1}}+A_2(x)\frac{\eta_{\mu\alpha}\eta_{\nu\beta}(x^\alpha x^\beta)^{d_2}}{l_2^{2d_2}}
  +\cdots
  +A_n(x)\frac{\eta_{\mu\alpha}\eta_{\nu\beta}(x^\alpha x^\beta)^{d_n}}{l_n^{2d_n}},
\end{align}
thus there are $n+1$ universal constants: $c, l_1,\cdots, l_n$, which will be all  set up to 1 for convenience.
 Let us consider  the following action\cite{l}
 \begin{align}\label{1}
   S=\int\sqrt { B_{\mu\nu} dx^\mu dx^\nu}=\int dt\sqrt{B_{00}+2B_{0i}v^i+B_{ij}v^iv^j},
 \end{align}
where $v^i=\frac{dx^i}{dt}, i=1,2,3$.
Thereby the  Lagrangian is given by
\begin{align}
 L(t,x^i,v^i)=\sqrt{-A_0+\sum_IA_It^{2d_I}-2\sum_IA_Iv^i(tx^i)^{d_I}+A_0v\cdot v+\sum_IA_I(x^ix^j)^{d_I}v^iv^j},
\end{align}
and the Euler-Lagrange equation reads
\begin{align}
  \frac{\partial L}{\partial x^i}=\frac{d}{dt}\frac{\partial L}{\partial v^i}=\frac{\partial}{\partial t}\frac{\partial L}{\partial v^i}+v^j\frac{\partial}{\partial x^j}\frac{\partial L}{\partial v^i}+\frac{dv^j}{dt}\frac{\partial^2 L}{\partial v^j v^i}.
\end{align}
 If a free particle of mass $m=1$ is assumed to be subject to this action, namely  the Euler-Lagrange equation  implies the acceleration has to vanish for the free particle (i.e. the law of inertia), we should have
 \begin{align}\label{2}
   \left\{
     \begin{array}{ll}
       \frac{\partial L}{\partial x^i}=\frac{\partial}{\partial t}\frac{\partial L}{\partial v^i}+v^j\frac{\partial}{\partial x^j}\frac{\partial L}{\partial v^i},\\
\det(\frac{\partial^2 L}{\partial v^j\partial v^i})\neq0.
     \end{array}
   \right.
 \end{align}

We observe that the unique promise of \eqref{2} being satisfied  may arrive when  taking $n=1, d_1=1$, then the first equation of \eqref{2} reduces to
\begin{footnotesize}\begin{align*}
 &-\partial_iA_0+\partial_iA_1t^2-2\partial_iA_1t\bm{x}\cdot \bm{v}+\partial_iA_0\bm{v}\cdot \bm{v}+\partial_iA_1(\bm{x}\cdot \bm{v})^2\\
 =&-2\partial_tA_1tx^i-2A_1x^i+2\partial_tA_0v^i+2\partial_tA_1x^i\bm{x}\cdot \bm{v}-2\bm{v}\cdot \nabla A_1tx^i+2\bm{v}\cdot \nabla A_0v^i+2\bm{v}\cdot \nabla A_1x^i\bm{x}\cdot \bm{v}+2A_1x^i\bm{v}\cdot \bm{v}\\
 &-\frac{(-\partial_t A_0+2A_1t+\partial_tA_1t^2-2A_1\bm{x}\cdot \bm{v}-2\partial_tA_1t\bm{x}\cdot \bm{v}+\partial_tA_0\bm{v}\cdot \bm{v}+\partial_tA_1(\bm{x}\cdot \bm{v})^2)(-2A_1tx^i+2A_0v^i+2A_1x^i\bm{x}\cdot \bm{v})}{2(-A_0+A_1t^2-2A_1t\bm{x}\cdot \bm{v}+A_0\bm{v}\cdot \bm{v}+A_1(\bm{x}\cdot \bm{v})^2)}\\
 &-\frac{(\bm{v}\cdot (-\nabla A_0+\nabla A_1t^2-2\nabla A_1 t\bm{x}\cdot \bm{v}+\nabla A_0 \bm{v}\cdot \bm{v}+\nabla A_1(\bm{x}\cdot \bm{v})^2-2\bm{v}A_1t+2\bm{v}A_1\bm{x}\cdot \bm{v}))(-2A_1tx^i+2A_0v^i+2A_1x^i\bm{x}\cdot \bm{v})}{2(-A_0+A_1t^2-2A_1t\bm{x}\cdot \bm{v}+A_0\bm{v}\cdot \bm{v}+A_1(\bm{x}\cdot \bm{v})^2)},
\end{align*}\end{footnotesize}
where the
following notations are employed
\begin{align*}
       \bm{x}&=x^1\frac{\partial}{\partial x^1}+x^2\frac{\partial}{\partial x^2}+x^3\frac{\partial}{\partial x^3},\\
        \bm{v}&=v^1\frac{\partial}{\partial x^1}+v^2\frac{\partial}{\partial x^2}+v^3\frac{\partial}{\partial x^3},\\
        \nabla A_{0/1}&=\partial_1 A_{0/1}\frac{\partial}{\partial x^1}+\partial_2 A_{0/1}\frac{\partial}{\partial x^2}+\partial_3 A_{0/1}\frac{\partial}{\partial x^3},
      \end{align*}
and $\bm{x}\cdot\bm{v}=\sum_ix^iv^i$, $\bm{v}\cdot\bm{v}=\sum_iv^iv^i$, $\bm{v}\cdot \nabla A_{0/1}=\sum_ix^i\partial_iA_{0/1}$.

Comparing the monomials of the both sides of the equality with the same type, we derive  the following 1-order partial differential equations
 \begin{align}
   \left\{
    \begin{array}{ll}
      \partial_i A_0=2A_1x^i,\\
\partial_t A_0=-2A_1t,\\
\partial_i A_1A_0=4A_1^2x^i,\\
\partial_t A_1A_0=-4A_1^2t,
    \end{array}
  \right.
 \end{align}
 thus
 \begin{align}
   \left\{
    \begin{array}{ll}
\partial_\mu A_0=2A_1\eta_{\mu\nu}x^\nu,\\
\partial_\mu A_1A_0=4A_1^2\eta_{\mu\nu}x^\nu.
    \end{array}
  \right.
 \end{align}
Obviously, if $A_0$ or $A_1$ is constant, then $A_1$ must vanish and $A_0$ is constant,  so everything essentially goes back to the classical theory when  $B=\eta$ up to a positive constant. The non-trivial solutions are given by
\begin{align}
  A_0=\frac{a}{b+\eta_{\mu\nu}x^\mu x^\nu},A_1=-\frac{a}{(b+\eta_{\mu\nu}x^\mu x^\nu)^2},
\end{align}
with two dimensionless constants $a,b$.  To check the second condition in \eqref{2}, we only need to show that under the limit $l_1\rightarrow\infty$, which is straightforward calculated
\begin{align*}
  \lim_{l_1\rightarrow\infty}\det(\frac{\partial^2 L}{\partial v^j\partial v^i})=-\frac{1}{|\frac{b}{a}(\bm{v}\cdot\bm{v}-1)|^{\frac{3}{2}}}\neq0.
\end{align*}

Moreover if we assume that $B$ has the same signature as $\eta$, then we has to require
\begin{align*}
 & B_{00}=-\frac{a(b+\bm{x}\cdot\bm{x})}{(b-t^2+\bm{x}\cdot\bm{x})^2}<0, \\
&\widetilde{B}_{11}=\frac{a(b+(x^2)^2+(x^3)^2)}{(b- t^2+\bm{x}\cdot\bm{x})(b+\bm{x}\cdot\bm{x})}>0,\\
&\det\left(
      \begin{array}{cc}
        \widetilde{B}_{11} & \widetilde{B}_{12} \\
        \widetilde{B}_{12} & \widetilde{B}_{22} \\
      \end{array}
    \right)=\frac{a^2(b+(x^3)^2)}{(b- t^2+ \bm{x}\cdot\bm{x})^2(b+\bm{x}\cdot\bm{x})}>0,\\
    &\det\left(
      \begin{array}{ccc}
        \widetilde{B}_{11} & \widetilde{B}_{12} & \widetilde{B}_{13}\\
        \widetilde{B}_{12} & \widetilde{B}_{22} & \widetilde{B}_{23}\\
       \widetilde{B}_{13}& \widetilde{B}_{23}& \widetilde{B}_{33}\\
      \end{array}
    \right)=\frac{a^3b}{(b- t^2+\bm{x}\cdot\bm{x})^3(b+\bm{x}\cdot\bm{x})}>0,
                                                                                 \end{align*}
where $\widetilde{B}_{ij}=B_{ij}-\frac{B_{0i}B_{0j}}{B_{00}}$, namely the following conditions should be imposed
\begin{align}\label{3}
a>0, b>0,\textrm{ and }b-t^2+\bm{x}\cdot\bm{x}>0.
\end{align}
or
\begin{align}\label{4}
a<0, b<0, b+\bm{x}\cdot\bm{x}<0.
\end{align}
For example, let $a=b=1$, and let the \emph{short distance approximation} $\frac{x^i}{l_1}\ll1$ for $i=1,2,3$ be adopted, then  the action \eqref{1} for a free particle  is simplified to be integrated out
\begin{align}\label{a}
S=&\int\sqrt{|B_{00}+(v^1)^2B_{11}+(v^2)^2B_{22}+(v^3)^2B_{33}|}dt\nonumber\\
=&\int\sqrt{\frac{1}{(1-t^2)^2}-\frac{\bm{v}\cdot\bm{v}}{1-t^2}}dt\nonumber\\
=&\frac{1}{2}\ln\frac{\sqrt{1-\bm{v}\cdot\bm{v}(1-t^2)}+t}{\sqrt{1-\bm{v}\cdot\bm{v}(1-t^2)}-t}-\sqrt{\bm{v}\cdot\bm{v}}\ln[\sqrt{\bm{v}\cdot\bm{v}}t+\sqrt{1-\bm{v}\cdot\bm{v}(1-t^2)}].
\end{align}

A remarkable perception is that $(A)dS_4$ geometry may emerge from the two-parameter dynamics of free particle in flat spacetime.  This picture exhibits some phenomenon of \emph{dynamics/geometry correspondence}.  $(A)dS_4$
is defined by a  hypersurface in 5-dimensional  space $\mathbb{R}^5$ with the Minkowski  metric $\eta^{(5)}=\textrm{diag}(-1,1,1,1,1)
$ (or the metric $\widetilde{\eta}^{(5)}=\textrm{diag}(-1,-1,1,1,1)
$ with  signature 1) via  the following equation\cite{ll}
\begin{align*}
    -T^2+X^2+Y^2+Z^2+bW^2=1 ( b>0),
\end{align*}
or\begin{align*}
  -T^2-bW^2+X^2+Y^2+Z^2=-1 ( b>0, T<1).
\end{align*}
Define the following coordinates which cover the half domain $\{W>0\}$ or $\{W<0\}$ in $(A)dS_4$
\begin{align}\label{t}
  x^0=\frac{T}{W},x^1=\frac{X}{W}, x^2=\frac{Y}{W},x^3=\frac{Z}{W},
\end{align}
Then the induced metric on  this hypersurface is given by in terms of the coordinate system $\{x^0,x^1,x^2,x^3\}$
\begin{align}\label{c}
  g_{\mu\nu}=\frac{b\eta_{\mu\nu}}{b+\eta_{\alpha\beta}x^\alpha x^\beta}-\frac{b\eta_{\mu\alpha}\eta_{\nu\beta}x^\alpha x^\beta}{(b+\eta_{\alpha\beta}x^\alpha x^\beta)^2},
\end{align}
or\begin{align}
  g_{\mu\nu}=\frac{b\eta_{\mu\nu}}{b-\eta_{\alpha\beta}x^\alpha x^\beta}+\frac{b\eta_{\mu\alpha}\eta_{\nu\beta}x^\alpha x^\beta}{(b-\eta_{\alpha\beta}x^\alpha x^\beta)^2},
\end{align}
which  exactly coincides with our quadratic form $B$ with condition \eqref{3} or \eqref{4} up to an insignificant constant conformal factor.

From this viewpoint, we
immediately conclude that the coordinate transformations preserve $B$ form the group $O(1,4)$ or $O(2,3)$ which contains Lorentz group $O(1,3)$ as a subgroup preserving the pair $(\eta,B)$, thus preserving the inertial motion.  By decomposing a matrix belogs to the group $O(1,4)$ or $O(2,3)$ as
\begin{align*}
  \lambda\left(
    \begin{array}{cc}
      N& P\\
      \mp\frac{P^T\eta N}{\sqrt{1\mp\eta_{\mu\nu}P^\mu P^\nu}} & \sqrt{1\mp\eta_{\mu\nu}P^\mu P^\nu}\\
    \end{array}
  \right),
\end{align*}
with matrices $N=(N_{\mu\nu})$ and $P=(P^0,P^1,P^2,P^3)^T$ satisfing the relation
\begin{align}
                                                                          N^T\eta N=\eta+\frac{N^T\eta PP^T\eta N}{\mp1+\eta_{\mu\nu}P^\mu P^\nu},
                                                                         \end{align}
                                                                         where $\mp$ correspond $O(1,4)$ and  $O(2,3)$ respectively, and $\lambda$ is fixed to 1 or $-$1, we can explicitly determine these coordinate transformations as fractional linear transformations\cite{x,y}
\begin{align}\label{b}
 x^\mu\mapsto\frac{N_{\mu\nu} x^\nu+\sqrt{b}P^\mu}{\mp\frac{\eta_{\alpha\beta}N_{\beta\gamma}P^\alpha x^\gamma}{\sqrt{1\mp\eta_{\mu\nu}P^\mu P^\nu}}+\sqrt{b}\sqrt{1\mp\eta_{\mu\nu}P^\mu P^\nu}},
\end{align}
which come back to Poinc\'{a}re transformations when the parameter $l_1$ tends to infinity.
Since the action (2) is invariant under these  transformations, there are corresponding conserved charges
for a free particle,  which can be given  via Norther method. Non-trivial charges would reflect the dynamical (not geometrical) symmetries.

\section{Unruh Effects Within the Context of Two-Parameter Dynamics}
A Klein-Gordon-type  equation that governs scalar field $\Phi$ with mass $m$ under the context of two-parameters dynamics is presented as
\begin{align}\label{k}
 \frac{1}{\sqrt{|\det (B_{\mu\nu})|}}\partial_\mu(\sqrt{|\det (B_{\mu\nu})|}B^{\mu\nu}\partial_\nu\Phi)-(m^2+\frac{\xi}{l_1^2})\Phi=0
\end{align}
for a dimensionless constant $\xi$,  which is invariant under the transformations \eqref{b}. Here the quadratic form $B$ is taking the form in \eqref{c} with $b=1$, and $(B^{\mu\nu})=(\frac{\eta^{\mu\nu}-x^\mu x^\nu}{1-t^2+\bm{x}\cdot\bm{x}})$ is the inverse of $B$. Assume that the scalar field is distributed on a small spatial domain,  thus the short distance approximation is valid. Changing variables as
\begin{equation}
\left\{
  \begin{array}{ll}
   \tilde{x}^0=\frac{1}{2}\ln\frac{1+t}{1-t},\\
\tilde{x}^i=x^i,i=1,2,3,
  \end{array}
\right.
\end{equation}
the equation \eqref{k} reduces to
\begin{equation}
  (\eta^{\mu\nu}\tilde{\partial}_\mu\tilde{\partial}_\nu+\tilde{\partial}_0\ln\sigma(\tilde{x}_0)\tilde{\partial}_0-\frac{m^2+\xi}{\sigma(\tilde{x}^0)})\Phi=0,
\end{equation}
where
\begin{equation}
\sigma(\tilde{x}_0)=1-(\frac{e^{2\tilde{x}_0}-1}{e^{2\tilde{x}_0}+1})^2.
\end{equation}

Taking ansatz $\Phi_{\bm{k}}=\chi_{\bm{k}}(\tilde{x}^0)e^{i\bm{k}\cdot\bm{x}}$, we obtain an ordinary differential equation satisfied by the coefficient $\chi_{\bm{k}}$
\begin{equation}\label{eq:q}
  \ddot{\chi}_{\bm{k}}+F(\tilde{x}^0)\dot{\chi}_{\bm{k}}+G(\tilde{x}^0)\chi_{\bm{k}}=0,
\end{equation}
where
\begin{align*}
 F(\tilde{x}^0)&=2\sqrt{1-\sigma(\tilde{x}_0)}=2\frac{e^{2\tilde{x}^0}-1}{e^{2\tilde{x}^0}+1},\\
  G(\tilde{x}^0)&=\frac{m^2+\xi}{\sigma(\tilde{x}^0)}+\bm{k}\cdot\bm{k}.
\end{align*}
We may introduce a family of solutions $\chi_{\bm{k}}=e^{i\frac{s(\tilde{x}^0)}{\lambda}}$ with
a parameter $\lambda$, and expand the function $s$ in terms of  $\lambda$ as $s=s_0+i\lambda s_1+(i\lambda)^2s_2+\cdots$. By considering the power of $\lambda$, we have
\begin{eqnarray*}
 (\frac{ds_0}{d\tilde{x}^0})^2=w^2(\tilde{x}^0),\\
\frac{d^2s_0}{d(\tilde{x}^0)^2}-2\frac{ds_0}{d\tilde{x}^0}\frac{ds_1}{d\tilde{x}^0}+F(\tilde{x}^0)\frac{ds_0}{d\tilde{x}^0}=0,\\
\frac{d^2s_1}{d(\tilde{x}^0)^2}-(\frac{ds_1}{d\tilde{x}^0})^2-2\frac{ds_0}{d\tilde{x}^0}\frac{ds_2}{d\tilde{x}^0}+F(\tilde{x}^0)\frac{ds_1}{d\tilde{x}^0}=0,\\
\cdots\cdots\ \ \ \ \ \ \ \ \ \ \ \ \nonumber
\end{eqnarray*}
From these recurrence relations we  get
\begin{align*}
  s_1=\ln\sqrt{w}+\int\frac{F(\tilde{x}^0)}{2}d\tilde{x}^0,\\
s_2=\frac{1}{4w^2}\frac{dw}{d\tilde{x}^0}+\frac{1}{8}\int d\tilde{x}^0[\frac{1}{w^3}(\frac{dw}{d\tilde{x}^0})^2+\frac{F^2}{w}+\frac{2}{w}\frac{dF}{d\tilde{x}^0}],\\
\cdots\cdots\ \ \ \ \ \ \ \ \ \ \ \ \nonumber
\end{align*}
Substituting   the first order solution
\begin{align}
  \chi_{\bm{k}}&=\frac{1}{\sqrt{w}}e^{-\int\frac{F(\tilde{x}^0)}{2}d\tilde{x}^0 }e^{\pm i\int w(\tilde{x}^0)d\tilde{x}^0}\nonumber\\
&=\frac{e^{\frac{\tilde{x}^0}{l}}}{1+e^{\frac{2\tilde{x}^0}{l}}}\frac{1}{\sqrt{w}}e^{\pm i\int w(\tilde{x}^0)d\tilde{x}^0}
\end{align}
into \eqref{eq:q} gives rise to the equation controls  $w$
\begin{equation*}
  w^2=G(\tilde{x}^0)-\frac{1}{2}\{\int wd\tilde{x}^0;\tilde{x}^0\}_S-\frac{1}{2}\frac{d F}{d\tilde{x}^0}-\frac{F^2}{4},
\end{equation*}
where $\{;\}_S$ denotes the
Schwartz derivative defined by
\begin{equation*}
 \{\int wd\tilde{x}^0;\tilde{x}^0\}_S=\frac{1}{w}\frac{d^2w}{d(\tilde{x}^0)^2}-\frac{3}{2w^2}(\frac{dw}{d\tilde{x}^0})^2.
\end{equation*}
The lowest order approximation is given by
\begin{equation}
  w^{(0)}=\pm\sqrt{\frac{(m^2+\xi)(e^{2\tilde{x}^0}+1)^2}{4e^{2\tilde{x}^0}}+\bm{k}\cdot\bm{k}-1},
\end{equation}
and then by iteration  we can get other higher order
approximate solutions.
In particular, when $m=\xi=0$ we have the exact solution
\begin{equation}\label{eq:a}
  \chi_{\bm{k}}=\frac{e^{\tilde{x}^0}}{1+e^{2\tilde{x}^0}}(\bm{k}\cdot\bm{k}-1)^{-\frac{1}{4}}e^{\pm i\sqrt{\bm{k}\cdot\bm{k}-1}\tilde{x}^0}.
\end{equation}

 Let us consider a detector that is  in the ground state with energy $E_0$ at initial time. If it detects a particle, it will transit to an excited state with energy $E>E_0$ and meanwhile the field $\Phi$ will transit from the vacuum $|0\rangle$ to a certain  excited state $|T\rangle$. Roughly speaking, the transition amplitude $\mathbb{A}$ in this process can be calculated as\cite{BD}
\begin{equation}
 \mathbb{A}\sim\int_{-\infty}^{+\infty}\langle T|\Phi(x)|0\rangle e^{iS[x](E-E_0)}dS
\end{equation}
where $x$ denotes the worldline of the detector and $S[x]$ stands for the action of the detector. The field $\Phi$ can be expanded in terms of  excited modes as
 \begin{equation}
  \Phi=\int d^3\bm{k}a_{\bm{k}}\psi_{\bm{k}}+\int d^3\bm{k}a^\dag_{\bm{k}}\psi^*_{\bm{k}}=\int d^3\bm{k}a_{\bm{k}}e^{i\bm{k}\cdot\bm{x}}\chi_{\bm{k}}+\int d^3\bm{k}a^\dag_{\bm{k}}e^{-i\bm{k}\cdot\bm{x}}\chi^*_{\bm{k}},
\end{equation}
where $a_{\bm{k}}$ and $a^\dag_{\bm{k}}$ are viewed respectively as annihilation and creation operators after quantization. Therefore we have
\begin{equation}\label{eq:2}
 \mathbb{A}\sim\int_{-\infty}^{+\infty}e^{-i\bm{k}\cdot\bm{x}}\chi^*_{\bm{k}}e^{iS(E-E_0)}dS,
\end{equation}
where $\bm{k}$ is the 3-momentum of the final state $|T\rangle$. Suppose the detector  uniformly moves in a constant speed $\bm{v}$, namely its action $S$ is given by \eqref{1}.  In particular, for a static detector we have $S\sim \widetilde{x}^0$(see \eqref{a}), hence
\begin{align}
  \mathbb{A}\sim(\bm{k}\cdot\bm{k}-1)^{-\frac{1}{4}}\int_{-\infty}^{+\infty}\frac{e^{\tilde{x}^0}}{1+e^{2\tilde{x}^0}}e^{i\tilde{x}^0(E-E_0\pm\sqrt{\bm{k}\cdot\bm{k}-1})}d\tilde{x}^0,
\end{align}
thus the Fourier transformation of the function $\frac{e^{\tilde{x}^0}}{1+e^{2\tilde{x}^0}}$.  Consequently, we arrive at
\begin{equation}\label{eq:d}
\mathbb{A}\sim\frac{\pi(\bm{k}\cdot\bm{k}-1)^{-\frac{1}{4}} }{e^{\frac{\pi }{2}(E-E_0\pm\sqrt{\bm{k}\cdot\bm{k}-1})}+e^{-\frac{\pi }{2}(E-E_0\pm\sqrt{\bm{k}\cdot\bm{k}-1})}}.
\end{equation}
However, we cannot take the minus sign in \eqref{eq:d} because
 the probability amplitude of detecting a particle with energy $E_p=\sqrt{(E-E_0)^2+1}$ is given by
$\mathbb{A}\sim\int_{-\infty}^{+\infty}\frac{e^{\tilde{x}^0}}{1+e^{2\tilde{x}^0}}d\tilde{x}^0=\pi$, which is not acceptable physically since it tends to infinity when the parameter $l_1$ tends to infinity. For the plus sign it is easy to see $\lim_{l_1\rightarrow+\infty}\mathbb{A}\rightarrow 0$ due to the energy of excited state  being higher than that of ground state.

\section{Symmetry  Breaking}

The Lie bracket among the basis $\{M_{AB}=-M_{BA}, A,B=0,\cdots,4\}$ of Lie algebra $\mathfrak{o}(1,4)$ is given by
\begin{align*}
  [M_{AB},M_{CD}]=\eta^{(5)}_{AD}M_{BC}+\eta^{(5)}_{BC}M_{AD}-\eta^{(5)}_{AC}M_{BD}-\eta^{(5)}_{BD}M_{AC}.
\end{align*}
Let $J_\mu=\frac{M_{\mu4}}{l_1}, \mu=0,\cdots,3$, then
\begin{align*}
  [J_\mu,J_\nu]&=-\frac{M_{\mu\nu}}{l_1^2},\\
  [J_\mu,M_{\alpha\beta}]&=\eta_{\mu\alpha}J_\beta-\eta_{\mu\beta}J_\alpha,\\
  [M_{\mu\nu},M_{\alpha\beta}]&=\eta_{\mu\beta}M_{\nu\alpha}+\eta_{\nu\alpha}M_{\mu\beta}-\eta_{\mu\alpha}M_{\nu\beta}-\eta_{\nu\beta}M_{\mu\alpha}.
\end{align*}
These relations can be realized via  the following differential operators
 \begin{align*}
       J_\mu&=\partial_\mu+\frac{\eta_{\mu\alpha}x^\alpha x^\nu\partial_\nu}{l_1^2},\\
       M_{\mu\nu}&=\eta_{\mu\alpha}x^\alpha\partial_\nu-\eta_{\nu\alpha}x^\alpha\partial_\mu=\eta_{\mu\alpha}x^\alpha J_\nu-\eta_{\nu\alpha}x^\alpha J_\mu.
                                                                          \end{align*}
Let us introduce the following symbols
\begin{align*}
    K_\mathfrak{i}^{\pm}&=\frac{1}{\sqrt2} (M_{0i}\pm M_{1i}), \mathfrak{i}=2,3\\
    F_{i}^{\pm}&=\frac{1}{\sqrt2} (\frac{M_{0i}}{l_1}\pm J_{i}),i=1,2,3,\\
    L_{i}&=\frac{1}{2}\epsilon_{ijk}M_{jk}, i,j,k=1,2,3,\\P^{\pm}&= \frac{1}{\sqrt2} (J_0\pm J_1),
      R=M_{01}, T=M_{23}.
                                   \end{align*}
The maximal Lie subalgebras of rank 7 in $\mathfrak{o}(1,4)$ are exhibited in the following list
$$\begin{tabular}{l|c|c}
  \hline
  & Generators & Algebraic Relations \\
    \hline
Type I & $\{K_1^{\pm},K_2^{\pm},J_2, J_3, P^{\pm},R,T\}$ & \begin{tabular}{c}
                                           $[K_\mathfrak{i}^{\pm},K_\mathfrak{j}^{\pm}] =0,[J_\mathfrak{i},J_\mathfrak{j}]=\epsilon_{\mathfrak{ij}}\frac{T}{l_1^2}, [K_\mathfrak{i}^{\pm},J_\mathfrak{j}]=\delta_{\mathfrak{ij}}P^{\pm}$,\\
  $[K_\mathfrak{i}^{\pm},P^{\pm}]=0,[K_\mathfrak{i}^{\pm},R]=-K_\mathfrak{i}^{\pm}, [K_\mathfrak{i}^{\pm},T]=\epsilon_{\mathfrak{ij}}K_\mathfrak{j}^{\pm}$,\\
  $[J_\mathfrak{i},P^{\pm}]=\frac{K_\mathfrak{i}^{\pm}}{l_1^2}, [J_\mathfrak{i},R]=0,[J_\mathfrak{i},T]=\epsilon_{\mathfrak{ij}}J_j$.\\
  $[P^{\pm},R]=\mp P^{\pm},[P^{\pm},T]=0,,[R,T]=0$,
                                           \end{tabular}
 \\ \hline
Type II & $\{F_1^{\pm},F_2^{\pm},F_3^{\pm},L_1,L_2,L_3,J_0\}$ & \begin{tabular}{c}
                                           $[F_{i}^{\pm},F_{j}^{\pm}]=0, [L_{i},L_{j}]=-\epsilon_{ijk}L_k, [F_{i}^{\pm},L_{j}]=-\epsilon_{ijk}F_k^{\pm}$,\\
                                           $[F_i^{\pm},J_0]=\pm\frac{F_i^{\pm}}{l_1^2},[L_i,J_0] = 0$.
                                           \end{tabular} \\
  \hline
\end{tabular}$$
The little groups in $O(1,4)$ corresponding to these two types of Lie subalgebras are denoted by $\mathcal{G}_1$ and $\mathcal{G}_2$ respectively, which exactly coincide with the groups $ISIM(2)$\footnote{In some literatures\cite{p}, $ISIM(2)$ means an 8-dimensional maximal subgroup of the Poincar\'{e} group generated by $\{K_1^{\pm},K_2^{\pm},J_1, J_2, P^+,P^-,R,T\}$.}and $O(3)\ltimes \mathbb{T}$ ($\mathbb{T}$ denoting the 4-dimensional translation group) respectively when the parameter $l_1$ tends to infinity. Some subgroups of $\mathcal{G}_1$ and $\mathcal{G}_2$ are listed as follows:
$$\begin{tabular}{l|c|c}
  \hline
   & Subgroups& Generators of Lie Algebras \\
    \hline
  $\mathcal{G}_1$&  \begin{tabular}{c}
                      $\mathcal{H}_1$\\
                  $\mathcal{H}_2$\\
                  $\mathcal{H}_3$\\
                     $\mathcal{H}_4$\\
                  $\mathcal{H}_5$\\
  $\mathcal{H}_6$\\
  $\mathcal{H}_7$\\
  $\mathcal{H}_8$
                    \end{tabular}&\begin{tabular}{c}
                      $\{K_1^{\pm},K_2^{\pm},P\}$\\
                  $\{K_1^{\pm},K_2^{\pm},P,R\}$\\
                  $\{K_1^{\pm},K_2^{\pm},P,T\}$\\
                     $\{K_1^{\pm},K_2^{\pm},P,R,T\}$\\
                  $\{P,R,T\}$\\
  $\{J_2,J_3,T\}$\\
                        $\{J_2,J_3,R, T\}$\\
                        $\{K_1^{\pm},K_2^{\pm},P, J_2,J_3, T\}$
                    \end{tabular}\\
  \hline
   $\mathcal{G}_2$&\begin{tabular}{c}
                      $\mathcal{K}_1$\\
                  $\mathcal{K}_2$\\
                  $\mathcal{K}_3$\\
                     $\mathcal{K}_4$\\
                     $\mathcal{K}_5$
                    \end{tabular}&\begin{tabular}{c}
                      $\{F_1^{\pm},F_2^{\pm},F_3^{\pm}\}$\\
                  $\{L_1,L_2,L_3\}$\\
                  $\{F_1^{\pm},F_2^{\pm},F_3^{\pm},J_0\}$\\
                     $\{L_1,L_2,L_3,J_0\}$\\
                     $\{F_1^{\pm},F_2^{\pm},F_3^{\pm},L_1,L_2,L_3\}$
                    \end{tabular}\\
    \hline
\end{tabular}$$

To construct an action whose symmetry group breaks into the little group $\mathcal{G}_1$ or $\mathcal{G}_2$ or one of  their subgroups, we  need find some invariant tensors under these groups.
\paragraph{Example} 1. For the subgroup $\mathcal{H}_1$ or $\mathcal{H}_3$, by  taking the following matrix representations of generators
\begin{align*}
  K_2^+&=\frac{1}{\sqrt2}\left(
          \begin{array}{ccccc}
            0 & 0 & 1 & 0 & 0 \\
           0 & 0 & 1 & 0 & 0 \\
            1& -1 & 0 & 0 & 0 \\
           0 & 0 & 0 & 0 & 0 \\
           0 & 0 & 0 & 0 & 0 \\
          \end{array}
        \right), K_3^+=\frac{1}{\sqrt2}\left(
          \begin{array}{ccccc}
            0 & 0 & 0 & 1 & 0 \\
           0 & 0 & 0 & 1 & 0 \\
            0& 0 & 0 & 0 & 0 \\
           1 & -1 & 0 & 0 & 0 \\
           0 & 0 & 0 & 0 & 0 \\
          \end{array}
        \right),\\
        P^+&=\frac{1}{l_1}\left(
          \begin{array}{ccccc}
            0 & 0 & 0 & 0 & 1 \\
           0 & 0 & 0 & 0 & 1 \\
            0& 0 & 0 & 0 & 0 \\
           0 & 0 & 0 & 0 & 0 \\
           1& -1 & 0 & 0 & 0 \\
          \end{array}
        \right),T=\left(
          \begin{array}{ccccc}
            0 & 0 & 0 & 0 & 0 \\
           0 & 0 & 0 & 0 & 0 \\
            0& 0 & 0 & 1 & 0 \\
           0 & 0 & -1 & 0 & 0 \\
           0& 0 & 0 & 0 & 0 \\
          \end{array}
        \right),
\end{align*}
we find a second-order non-degenerate symmetric invariant tensor with respect to $\mathcal{H}_1$ or $\mathcal{H}_3$
\begin{align}
 C=\left(
     \begin{array}{ccccc}
       \textsf{ a} & \textsf{b} & 0 & 0 & 0 \\
           \textsf{b} & 2\textsf{b}-\textsf{a} & 0 & 0 & 0 \\
            0& 0 & \textsf{b}-\textsf{a }& 0 & 0 \\
           0 & 0 & 0 & \textsf{b}-\textsf{a} & 0 \\
           0& 0 & 0 & 0 & \textsf{b}-\textsf{a} \\
     \end{array}
   \right)
\end{align}
with two constants \textsf{a}$\neq$\textsf{b}, thus a quadratic form
\begin{align}
  C=\textsf{a}dT^2+2\textsf{b}dTdX+(2\textsf{b}-\textsf{a})dX^2+(\textsf{b}-\textsf{a})dY^2+(\textsf{b}-\textsf{a})dZ^2+(\textsf{b}-\textsf{a})dW^2.
\end{align}
Then the coordinate transformations \eqref{t} give rises to an induced quadratic form
\begin{align}
 C=& C_{\mu\nu}dx^\mu dx^\nu\nonumber\\
=&\frac{1}{(1+\eta_{\mu\nu}x^\mu x^\nu)^2}\{[\textsf{b}\frac{(1+\bm{x}\cdot\bm{x}+x^0x^1)^2}{1+\eta_{\mu\nu}x^\mu x^\nu}-(\textsf{b}-\textsf{a})(1+\bm{x}\cdot\bm{x})](dx^0)^2\nonumber\\&+
  2[\textsf{b}\frac{(1+\bm{x}\cdot\bm{x}+x^0x^1)(1+\eta_{\mu\nu}x^\mu x^\nu-(x^1)^2-x^0x^1)}{1+\eta_{\mu\nu}x^\mu x^\nu}+(\textsf{b}-\textsf{a})x^0x^1]dx^0dx^1\nonumber\\&+
  2[-\textsf{b}\frac{(1+\bm{x}\cdot\bm{x}+x^0x^1)(x^0x^2+x^1x^2)}{1+\eta_{\mu\nu}x^\mu x^\nu}+(\textsf{b}-\textsf{a})x^0x^2]dx^0dx^2\nonumber\\
  &+2[-\textsf{b}\frac{(1+\bm{x}\cdot\bm{x}+x^0x^1)(x^0x^3+x^1x^3)}{1+\eta_{\mu\nu}x^\mu x^\nu}+(\textsf{b}-\textsf{a})x^0x^3]dx^0dx^3\nonumber\\
  &+[\textsf{b}\frac{(1+\eta_{\mu\nu}x^\mu x^\nu-(x^1)^2-x^0x^1)^2}{1+\eta_{\mu\nu}x^\mu x^\nu}+(\textsf{b}-\textsf{a})(1+\eta_{\mu\nu}x^\mu x^\nu-(x^1)^2)](dx^1)^2\nonumber\\&+
  2[-\textsf{b}\frac{(1+\eta_{\mu\nu}x^\mu x^\nu-(x^1)^2-x^0x^1)(x^0x^2+x^1x^2)}{1+\eta_{\mu\nu}x^\mu x^\nu}-(\textsf{b}-\textsf{a})x^1x^2]dx^1dx^2\nonumber\\&+
  2[-\textsf{b}\frac{(1+\eta_{\mu\nu}x^\mu x^\nu-(x^1)^2-x^0x^1)(x^0x^3+x^1x^3)}{1+\eta_{\mu\nu}x^\mu x^\nu}-(\textsf{b}-\textsf{a})x^1x^3]dx^1dx^3\nonumber\\
  &+[\textsf{b}\frac{(x^0x^2+x^1x^2)^2}{1+\eta_{\mu\nu}x^\mu x^\nu)}+(\textsf{b}-\textsf{a})(1+\eta_{\mu\nu}x^\mu x^\nu-(x^2)^2)](dx^2)^2\nonumber\\&+
  2[\textsf{b}\frac{x^2x^3(x^0+x^1)^2}{1+\eta_{\mu\nu}x^\mu x^\nu}-(\textsf{b}-\textsf{a})x^2x^3]dx^2dx^3\nonumber\\&+[\textsf{b}\frac{(x^0x^3+x^1x^3)^2}{1+\eta_{\mu\nu}x^\mu x^\nu}+(\textsf{b}-\textsf{a})(1+\eta_{\mu\nu}x^\mu x^\nu-(x^3)^2)](dx^3)^2\}.
\end{align}
Hence the invariant action can be chosen as 
\begin{align}
  S=\int\sqrt{C_{\mu\nu} dx^\mu dx^\nu}.
\end{align}
 
2. Similarly for the subgroup $\mathcal{H}_1$, we also have a second-order anti-symmetric tensor
\begin{align}
  D=\left(
      \begin{array}{ccccc}
        0 & 0 &\textsf{a} & \textsf{b} & \textsf{c} \\
        0 & 0 & \textsf{a} & \textsf{b} &\textsf{c}\\
        -\textsf{a }& -\textsf{a} & 0 & 0 & 0 \\
        -\textsf{b }& -\textsf{b} & 0 & 0 & 0 \\
       -\textsf{c }& -\textsf{c} & 0 & 0 & 0 \\
      \end{array}
    \right)
\end{align}
with three constants $\textsf{a},\textsf{b},\textsf{c}$,
thus a 2-form
\begin{align}
  D=\textsf{a}(dT+dX)\wedge dY+\textsf{b}(dT+dX)\wedge dZ+\textsf{c}(dT+dX)\wedge dW.
\end{align}
Therefore the following Yang-Mills-type action is $\mathcal{H}_1$-invariant 
\begin{align}
  S=\int d^4x D_{\mu\nu}D_{\alpha\beta}B^{\mu\alpha}B^{\nu\beta},
\end{align}
where the induced 2-form $D$ is given by 
\begin{align}
  D=&\frac{1}{2}D_{\mu\nu}dx^\mu\wedge dx^\nu\nonumber\\
=&\frac{1}{(1+\eta_{\mu\nu}x^\mu x^\nu)^2}[-(x^0+x^1)(\textsf{a}x^2+\textsf{b}x^3+\textsf{c})dx^0\wedge dx^1\nonumber\\&+(\textsf{a}(1+(x^1)^2+(x^3)^2+x^0x^1)-\textsf{b}x^2x^3-\textsf{c}x^2)dx^0\wedge dx^2\nonumber\\
&+(\textsf{b}(1+(x^1)^2+(x^2)^2+x^0x^1)-\textsf{a}x^2x^3-\textsf{c}x^3)dx^0\wedge dx^3\nonumber\\&+(\textsf{a}(1-(x^0)^2+(x^3)^2-x^0x^1)-\textsf{b}x^2x^3-\textsf{c}x^2)dx^1\wedge dx^2\nonumber\\&+(\textsf{b}(1-(x^0)^2+(x^2)^2-x^0x^1)-\textsf{a}x^2x^3-\textsf{c}x^3)dx^1\wedge dx^3\nonumber\\&+(x^0+x^1)(\textsf{a}x^3-\textsf{b}x^2)dx^2\wedge dx^3],
\end{align}which can be viewed as the strength of the field 
\begin{align}
 U=U_\mu dx^\mu=&\frac{x^0+x^1}{1+\eta_{\mu\nu}x^\mu x^\nu}[x^0(\textsf{a}x^2+\textsf{b}x^3+\textsf{c})dx^0-x^1(\textsf{a}x^2+\textsf{b}x^3+\textsf{c})dx^1\nonumber\\
&+(\textsf{a}(1+\eta_{\mu\nu}x^\mu x^\nu)-x^2(\textsf{a}x^2+\textsf{b}x^3+\textsf{c}))dx^2\nonumber\\
&+(\textsf{a}(1+\eta_{\mu\nu}x^\mu x^\nu)-x^3(\textsf{a}x^2+\textsf{b}x^3+\textsf{c}))dx^3].
\end{align}
Alterative choice is taking a  Born-Infeld-type action\cite{t}
\begin{align}
  S=\int d^4x\sqrt{|\det(B-DB^{-1}D)|}.
\end{align}

3. For the subgroup $\mathcal{K}_1$ whose generators are explicitly expressed  as  
\begin{align*}
  F_1^+&=\frac{1}{\sqrt2l_1}\left(
          \begin{array}{ccccc}
            0 & 1 & 0 & 0 & 0 \\
           1 & 0 & 0 & 0 & 1 \\
            0& 0 & 0 & 0 & 0 \\
           0 & 0 & 0 & 0 & 0 \\
           0 & -1 & 0 & 0 & 0 \\
          \end{array}
        \right), F_2^+=\frac{1}{\sqrt2l_1}\left(
          \begin{array}{ccccc}
            0 & 0 & 1 & 0 & 0 \\
           0 & 0 & 0 & 0 & 0 \\
            1& 0 & 0 & 0 & 1 \\
           0 & 0 & 0 & 0 & 0 \\
           0 & 0 & -1 & 0 & 0 \\
          \end{array}
        \right),
        F_3^+&=\frac{1}{\sqrt2l_1}\left(
          \begin{array}{ccccc}
            0 & 0 & 0 & 1 & 0 \\
           0 & 0 & 0 & 0 & 0 \\
            0& 0 & 0 & 0 & 0 \\
           1 & 0 & 0 & 0 & 1 \\
           0&0 & 0 & -1 & 0 \\
          \end{array}
        \right),
\end{align*}
there is an invariant vector 
\begin{align}
  V=(\textsf{a},0,0,0,-\textsf{a})^T
\end{align}
with a constant $\textsf{a}$.
Therefore we can consider a \emph{Finsler-type action}\cite{p,k} 
\begin{align}
  S=\int(B_{\mu\nu}dx^\mu dx^\nu)^{\frac{1-\delta}{2}}(V_\mu dx^\mu)^\delta
\end{align}
with   a constant $\delta\neq0,1$, thus the corresponding Lagrangian is given by
\begin{align}
 L(t,x^i,v^i)=(B_{00}+2B_{0i}v^i+B_{ij}v^iv^j)^{\frac{1-\delta}{2}}(V_0+V_iv^i)^\delta,
\end{align}
where 
\begin{align}
\left\{
  \begin{array}{ll}
    V_0 &= \textsf{a}(1+\eta_{\mu\nu}x^\mu x^\nu)^{-\frac{3}{2}}(1+\bm{x}\cdot\bm{x}-x^0),\\
  V_i&= \textsf{a}(1+\eta_{\mu\nu}x^\mu x^\nu)^{-\frac{3}{2}}x^i(1-x^0),i=1,2,3.
  \end{array}
\right.
\end{align}
 The pair $(B,V)$ is preserved by $\mathcal{K}_1$.

4. Since there are no new invariant tensors for the group $\mathcal{G}_1$ or $\mathcal{G}_2$, any local term appears 
in the Lagrangian enjoys the full  symmetry group $O(1,4)$,  and hence the symmetry
breaking effects are necessarily nonlocal\cite{v}. For the vectors $W=(\textsf{a},\textsf{a},0,0,0)^T$ and $V$, we have 
\begin{align*}
  K_{\mathfrak{i}}^+W&=J_{\mathfrak{i}}W=TW=PW=0, RW=W,\\
&F_{i}^+V=J_{i}V=0,J_0V=-V,
\end{align*}
namely the group $\mathcal{G}_1$ or $\mathcal{G}_2$ preserves the direction of the vector $W$ or $V$. Then we can write the following equation contains a nonlocal term for the scalar field $\Phi$ with mass $m$
\begin{align}
  \frac{1}{\sqrt{|\det B|}}\partial_\mu(\sqrt{|\det B|}B^{\mu\nu}(\partial_\nu+m^2\frac{W_\nu}{B^{\alpha\beta}W_\alpha\partial_\beta})\Phi)=0,
\end{align}where
\begin{align}
  W=W_\mu dx^\mu=\textsf{a}(1+\eta_{\mu\nu}x^\mu x^\nu)^{-\frac{3}{2}}[&(1+\eta_{\mu\nu}x^\mu x^\nu)d(x^0+x^1)\nonumber\\
&+(x^0+x^1)(x^0dx^0-x^1dx^1-x^2dx^2-x^3dx^3)].
\end{align}

 \end{document}